\documentclass[journal]{IEEEtran}

\usepackage{lineno,hyperref}
\usepackage{bm}
\usepackage{mathrsfs}
\usepackage{amsmath}
\usepackage{amssymb}
\usepackage{graphicx}
\usepackage{subfigure}
\usepackage{epstopdf}
\usepackage{xcolor}
\usepackage{wasysym}
\usepackage{footnote}
\usepackage{amssymb}
\usepackage{overpic}
\ifCLASSINFOpdf

\else

\fi

\begin{document}

\title{\Huge Study of Robust Sparsity-Aware RLS algorithms with Jointly-Optimized Parameters for Impulsive Noise Environments}

\author{Yi~Yu,~\IEEEmembership{Member,~IEEE},
        ~Lu~Lu,~\IEEEmembership{Member,~IEEE},
        ~Yuriy Zakharov,~\IEEEmembership{Senior Member,~IEEE},
        ~Rodrigo C. de Lamare,~\IEEEmembership{Senior Member,~IEEE},
        ~and
        ~Badong Chen,~\IEEEmembership{Senior Member,~IEEE}

        % <-this % stops a space
\thanks{This work was supported by National Science Foundation of P.R. China (Nos. 61901400 and 61901285) and Doctoral Research Fund of Southwest University of Science and Technology (No. 19zx7122). The work of Y. Zakharov was supported in part by the U.K. EPSRC through Grants EP/R003297/1 and EP/V009591/1. Corresponding author: Lu Lu. }

\thanks{Y. Yu is with School of Information Engineering, Robot Technology Used for Special Environment Key Laboratory of Sichuan Province, Southwest University of Science and Technology, Mianyang, 621010, China (e-mail: yuyi\_xyuan@163.com).}

\thanks{L. Lu is with School of Electronics and Information Engineering, Sichuan University, Chengdu, 610065, China (e-mail: lulu19900303@126.com).}

\thanks{Y. Zakharov is with the Department of Electronics, University of York, York YO10 5DD, U.K. (e-mail: yury.zakharov@york.ac.uk).}

\thanks{R. C. de Lamare is with CETUC, PUC-Rio, Rio de Janeiro 22451-900, Brazil. (e-mail: delamare@cetuc.puc-rio.br).}

\thanks{B. Chen is with Institute of Artificial Intelligence and Robotics, Xi'an Jiaotong University, Xi'an, 710049, Shaanxi Province, China. (e-mail: chenbd@mail.xjtu.edu.cn).}
 % stops a space
}

%\markboth{Journal of \LaTeX\ Class Files,~Vol.~14, No.~8, August~2015}%
%{Shell \MakeLowercase{\textit{et al.}}: Bare Demo of IEEEtran.cls for IEEE Journals}

\maketitle

\begin{abstract}
This paper proposes a unified sparsity-aware robust recursive
least-squares RLS (S-RRLS) algorithm for the identification of
sparse systems under impulsive noise.  The proposed algorithm
generalizes multiple algorithms only by replacing the specified
criterion of robustness and sparsity-aware penalty. Furthermore, by
jointly optimizing the forgetting factor and the sparsity penalty
parameter, we develop the jointly-optimized S-RRLS (JO-S-RRLS)
algorithm, which not only exhibits low misadjustment but also can
track well sudden changes of a sparse system. Simulations in
impulsive noise scenarios demonstrate that the proposed S-RRLS and
JO-S-RRLS algorithms outperform existing techniques.
\end{abstract}

% Note that keywords are not normally used for peerreview papers.

\IEEEpeerreviewmaketitle

\section{Introduction}

\IEEEPARstart{I}{n} realistic environments, in addition to Gaussian
noise, impulsive noise is often present such as in echo
cancellation, underwater acoustics, audio processing, and
communications~\cite{nikias1995signal,zimmermann2002analysis,yu2019dcd}.
Realizations of the impulsive noise are random and sparse in the
time domain, and typically have much higher amplitudes than Gaussian
noise. In these scenarios, the popular recursive least squares (RLS)
algorithm experiences significant performance deterioration.

Aiming at impulsive noise cases, various strategies have been
applied to develop robust adaptive algorithms
\cite{chan2004recursive,navia2012combination,zayyani2014continuous,lu2018recursive,lu2016improved,radmanesh2018recursive,chen2016generalized}, %,chen2017kernel}%,chen2017kernel};%,lee2009subband,pradhan2017improved}
\cite{jidf,spa,intadap,mbdf,jio,jiols,jiomimo,sjidf,ccmmwf,tds,mfdf,l1stap,mberdf,jio_lcmv,locsme,smtvb,ccmrls,dce,itic,jiostap,aifir,ccmmimo,vsscmv,bfidd,mbsic,wlmwf,bbprec,okspme,rdrcb,smce,armo,wljio,saap,vfap,saalt,mcg,sintprec,stmfdf,1bitidd,jpais,did,rrmber,memd,jiodf,baplnc,als,vssccm,doaalrd,jidfecho,dcg,rccm,ccmavf,mberrr,damdc,smjio,saabf,arh,lsomp,jrpaalt,smccm,vssccm2,vffccm,sor,aaidd,lrcc,kaesprit,lcdcd,smbeam,ccmjio,wlccm,dlmme,listmtc,smcg},
In particular, the recursive least M-estimate (RLM) algorithm was
proposed by introducing the M-estimator of the error
signal~\cite{chan2004recursive}; it not only shows good robustness
against impulsive noises, but also almost retains the fast
convergence which the RLS algorithm possesses in the Gaussian noise.
The correntropy defines similarity between two variables so that it
has the property to identify normal and abnormal
samples~\cite{chen2016generalized}. Based on this property, the
recursive maximum correntropy criterion (RMCC) algorithm was
presented~\cite{radmanesh2018recursive}, which is capable of dealing
with impulsive noises.

In studying adaptive algorithms, there is interest to exploit the system sparsity. A sparse vector signifies that it only has a few non-zero entries such as the impulse responses of propagation channels in underwater acoustic and terrestrial communications~\cite{radecki2002echo,schreiber1995advanced,zakharov2013dcd}. With different sparsity-aware penalties (e.g., the $l_1$-norm, $l_0$-norm, and logarithmic based penalties), several sparsity-aware variants of the standard RLS algorithm were reported~\cite{eksioglu2011rls,eksioglu2011sparsity}, which reduce the steady-state misadjustment. Likewise, in the impulsive noise environment, sparsity-aware robust RLS-type algorithms were proposed, such as the sparsity-aware RLM algorithm~\cite{pelekanakis2014adaptive} and the sparsity-aware RMCC algorithm~\cite{ma2017recursive}. However, it is worth noting that these algorithms have two drawbacks. First, they use the constant forgetting factor that controls the trade-off between the steady-state misadjustment and tracking capability for sudden changes of systems. Second, their performance depends on the sparsity-penalty parameter, which is usually chosen in a trial-and-error way, thus limiting their usefulness. In the literature, these two problems were separately considered for Gaussian noise scenarios. As such, variable forgetting factor (VFF)~\cite{paleologu2008robust,bhotto2012new,paleologu2014practical} and variable sparsity-penalty parameter (VSPP)~\cite{eksioglu2011rls} schemes have been developed. Nevertheless, they have not been considered jointly and for impulsive noise scenarios.
%Moreover, the VSPP scheme in~\cite{eksioglu2011rls} also requires the priori knowledge of the system sparsity.

In this paper, our contributions are as follows. 1) We propose a unified S-RRLS framework, which covers different algorithms by applying straightforwardly the specified robustness criterion and sparsity-aware penalty. 2) By decoupling the effect of forgetting factor and sparsity-penalty parameter from each other, we derive adaptive rules for adjusting online these two parameters, and therefore the resulting jointly-optimized S-RRLS (JO-S-RRLS) algorithm reaches a low steady-state misadjustment and good tracking capability in impulsive noise.

The remaining part of this paper is organized as follows. Section~II reviews the signal model and introduces the S-RRLS framework. In Section~III, we present the JO-S-RRLS algorithm. Section~IV shows simulations results. Conclusions are presented in Section~V.
\vspace{-3mm}
\section{Signal Model and S-RRLS Algorithm}
Consider a system identification problem with the input signal $x_k$ at time~$k$, then the output $d_k$ of the system is described by
\begin{align}
\label{001}
d_k = \bm x_k^\text{T} \bm w^o+v_k,
\end{align}
where $\bm w^o \in \mathbb{R}^{M\times1}$ is the impulse response of the sparse system to be identified, $\bm x_k=[x_k,x_{k-1},...,x_{k-M+1}]^\text{T} \in \mathbb{R}^{M\times1}$ is the input vector, and $v_k$ is the noise. An estimate of $\bm w^o$ at time~$k$, the vector of coefficients of adaptive filter, is denoted by $\bm w_k \in \mathbb{R}^{M\times1}$. For updating the filter coefficients in Gaussian noise, the sparsity-aware RLS algorithm that exploits the sparsity of $\bm w^o$ is popular. However, the surroundings noise may also include impulsive samples which deteriorate the algorithm performance. For such scenarios, we introduce the unified robust sparsity-aware minimization problem:
\begin{align}
\bm w_k = \arg \min\limits_{\bm w_k} \left\lbrace \sum\limits_{j=0}^k \lambda^{k-j} \varphi \left(d_j-\bm x_{j}^\text{T}\bm w_k\right) + \rho f(\bm w_k) \right\rbrace,
\label{002}
\end{align}
where $0<\lambda<1$ is the forgetting factor and $\varphi(\cdot): \mathbb{R} \rightarrow \mathbb{R}$ is a robustness function that can deal with impulsive noise. In~\eqref{002}, $f(\cdot): \mathbb{R}^{M\times1} \rightarrow \mathbb{R}$ is a general sparsity-aware penalty function, and $\rho>0$ is the sparsity-aware penalty parameter defining the weight of this penalty term. By solving the optimization in~\eqref{002}, we obtain the following normal equation:
\begin{align}
\bm R_k \bm w_k = \bm z_k- \rho \triangledown f(\bm w_k) ,
\label{003}
\end{align}
where
\begin{equation}
\begin{array}{rcl}
\begin{aligned}
\bm R_k = \sum\limits_{j=0}^k \lambda^{k-j} q_j \bm x_j \bm x_j^\text{T} = \lambda \bm R_{k-1}+q_k \bm x_k \bm x_k^\text{T} \\
\bm z_k = \sum\limits_{j=0}^k \lambda^{k-j} q_j d_j \bm x_j = \lambda \bm z_{k-1}+q_k d_n \bm x_k
\label{004}
\end{aligned}
\end{array}
\end{equation}
are the time-averaged autocorrelation matrix of $\bm x_k$ and the time-averaged crosscorrelation vector of $d_k$ and $\bm x_k$, respectively, and $\triangledown f(\bm w_k)$ is the subgradient of $f(\bm w_k)$ with respect to $\bm w_k$. Also, $q_k=\varphi'(\epsilon_k)/\epsilon_k$, where $\epsilon_k=d_k-\bm x_{k}^\text{T}\bm w_k$ is the \emph{a posteriori} error and $\varphi'(\epsilon_k)$ is the derivative of $\varphi(\epsilon_k)$.

Applying the matrix inversion lemma~\cite{sayed2003fundamentals}, we have
\begin{equation}
\label{005}
\begin{array}{rcl}
\begin{aligned}
\bm P_{k} \triangleq \bm R_{k}^{-1} = \frac{1}{\lambda} \left( \bm P_{k-1} - \bm K_{k} \bm x_{k}^T\bm P_{k-1} \right),
\end{aligned}
\end{array}
\end{equation}
where
\begin{equation}
\label{006}
\begin{array}{rcl}
\begin{aligned}
\bm K_{k} \triangleq \frac{q_k \bm P_{k-1}\bm x_{k}}{\lambda + q_k \bm x_{k}^T \bm P_{k-1} \bm x_{k}}
\end{aligned}
\end{array}
\end{equation}
is the Kalman gain vector, and $\bm P_{k}$ is initialized by $\bm P_{0}=\delta^{-1} \bm I_M$ with $\delta>0$ being the regularization parameter. Then, substituting~\eqref{005} and \eqref{006} into~\eqref{003}, the recursion for $\bm w_k$ is established:
\begin{equation}
\label{007}
\begin{array}{rcl}
\begin{aligned}
\bm w_k = \bm w_{k-1} + \bm K_{k} e_k - \rho \bm P_{k} [\triangledown f(\bm w_k) - \lambda \triangledown f(\bm w_{k-1})],
\end{aligned}
\end{array}
\end{equation}
where $e_k=d_k-\bm x_{k}^\text{T}\bm w_{k-1}$ is the \emph{a priori} error. Naturally, the above recursion is not realizable, since $q_k$ and $\triangledown f(\bm w_k)$ require knowing $\bm w_k$ at time~$k$ beforehand. To overcome this problem, it is assumed that $q_k$ and $\triangledown f(\bm w_k)$ do not change considerably at adjacent time. Hence, we can rewrite~\eqref{007} as
\begin{equation}
\label{008}
\begin{array}{rcl}
\begin{aligned}
\bm w_k = \bm w_{k-1} + \bm K_{k} e_k - \rho \bm P_{k} \triangledown f(\bm w_{k-1}),
\end{aligned}
\end{array}
\end{equation}
where $(1-\lambda)$ is absorbed into $\rho$ that is $\rho \leftarrow\rho(1-\lambda)$, and we can approximate $q_k$ as
 \begin{align}
 q_k \approx \varphi'(e_k)/e_k.
 \label{009}
 \end{align}
\vspace{-10mm}
\section{Proposed JO-S-RRLS Algorithm}
Clearly, the S-RRLS algorithm performance depends on the parameters $\lambda$ and $\rho$. Specifically, $\lambda$ governs the trade-off between steady-state error and tracking capability through the Kalman gain vector. The parameter~$\rho$ must be properly chosen to ensure that the S-RRLS algorithm fully exploits the system's sparsity and thus outperforms the RRLS algorithm. In this section, we design adaptive schemes for $\lambda$ and $\rho$, producing time-varying parameters $\lambda_k$ and $\rho_k$. For this purpose, we rearrange~\eqref{008} in two steps as
\begin{subequations} \label{eq:10}
    \begin{align}
    %   \begin{array}{rcl}
    %   \begin{aligned}
    \bm \psi_k &= \bm w_{k-1} + \bm K_{k} e_k,
    %    \end{aligned}
    %   \end{array}
    \label{eq:10a}\\
    \bm w_k &= \bm \psi_k - \rho_k \bm P_{k} \triangledown f(\bm \psi_{k}).\label{eq:10b}
    \end{align}
\end{subequations}
 The step~\eqref{eq:10a} plays the adaptive learning role of the RRLS algorithm; the step~\eqref{eq:10b} drives inactive coefficients in $\bm \psi_k$ to zero, thereby improving the performance of identifying sparse $\bm w^o$. Importantly, we also replace the original $\triangledown f(\bm w_{k-1})$ in \eqref{eq:10b} with $\triangledown f(\bm \psi_{k})$. In doing so, $\lambda_k$ and $\rho_k$ will be designed independently according to~\eqref{eq:10a} and~\eqref{eq:10b}, respectively.
\vspace{-7mm}
\subsection{Design of $\rho_k$}
By subtracting~\eqref{eq:10b} from $\bm w^o$, we obtain
\begin{equation}
\label{011}
\begin{array}{rcl}
\begin{aligned}
\widetilde{\bm w}_k = \widetilde{\bm \psi}_k + \rho_k \bm P_{k} \triangledown f(\bm \psi_{k}),
\end{aligned}
\end{array}
\end{equation}
where $\widetilde{\bm w}_k \triangleq \bm w^o - \bm w_k$ and $\widetilde{\bm \psi}_k \triangleq \bm w^o - \bm \psi_k$ represent the deviation vectors for the estimates $\bm w_k$ and $\bm \psi_k$. Taking the $l_2$-norm on both sides of~\eqref{011}, it is found that
\begin{equation}
\label{012}
\begin{array}{rcl}
\begin{aligned}
||\widetilde{\bm w}_k&||_2^2 = ||\widetilde{\bm \psi}_k||_2^2 + \\
& \underbrace{2\rho_k \widetilde{\bm \psi}_k^T \left( \bm P_{k} \triangledown f(\bm \psi_{k})\right) + \rho_k^2 ||\bm P_{k} \triangledown f(\bm \psi_{k})||_2^2}_{\bigtriangleup_k}.
\end{aligned}
\end{array}
\end{equation}

Minimizing $||\widetilde{\bm w}_k||_2^2$ with respect to $\rho_k$, the optimal $\rho_k$ is obtained as
\begin{equation}
\label{013}
\begin{array}{rcl}
\begin{aligned}
\rho_k^{\text{opt}} = \frac{[\bm \psi_k-\bm w^o]^T \left( \bm P_{k} \triangledown f(\bm \psi_{k})\right)}{||\bm P_{k} \triangledown f(\bm \psi_{k})||_2^2}.
\end{aligned}
\end{array}
\end{equation}
At time $k$, although we do not know the true $\bm w^o$ in~\eqref{013}, it can be approximated by the previous estimate $\bm w_{k-1}$ from~\eqref{eq:10b}. Thus, we modify~\eqref{013} to the following rule for choosing $\rho_k$:
\begin{equation}
\label{014}
\begin{array}{rcl}
\begin{aligned}
\hat{\rho}_k^{\text{opt}} = \max \left[ \frac{[\bm \psi_k-\bm w_{k-1}]^T \left( \bm P_{k} \triangledown f(\bm \psi_{k})\right)}{||\bm P_{k} \triangledown f(\bm \psi_{k})||_2^2},\;0\right].
\end{aligned}
\end{array}
\end{equation}
Also, since the estimate $\bm w_{k}$ at the early stage of the adaptation has a larger deviation as compared to $\bm w^o$, we enforce $\hat{\rho}_k^{\text{opt}}=0$ when $k \leq M/2$, to ensure the algorithm's convergence.
\subsection{Design of $\lambda_k$}
To derive the VFF scheme, we insert~\eqref{006} into~\eqref{eq:10a} to yield
\begin{equation}
\label{015}
\begin{array}{rcl}
\begin{aligned}
\bm \psi_k &= \bm w_{k-1} + \frac{q_k \bm P_{k-1}\bm x_{k}}{\lambda_k + \theta_k} e_k,
\end{aligned}
\end{array}
\end{equation}
where $\theta_k = q_k \bm x_{k}^T \bm P_{k-1} \bm x_{k}$. By defining the intermediate error $\xi_k=d_k-\bm x_{k}^\text{T} \bm \psi_k$, we can find from~\eqref{015} that
\begin{equation}
\label{016}
\begin{array}{rcl}
\begin{aligned}
\xi_k = \left( 1 - \frac{\theta_k}{\lambda_k + \theta_k}\right) e_k.
\end{aligned}
\end{array}
\end{equation}
Inspired by~\cite{paleologu2008robust}, we propose to impose the condition
\begin{equation}
\label{017}
\begin{array}{rcl}
\begin{aligned}
E\{\xi_k^2\}=\sigma_v^2
\end{aligned}
\end{array}
\end{equation}
on~\eqref{016}, which helps to recover the system noise in the intermediate error signal, where $E\{\cdot\}$ denotes the mathematical expectation. However, it is emphasized that $\sigma_v^2$ in~\eqref{017} denotes the variance of the noise excluding impulsive noise samples, due to the fact that the negative influence of impulsive noises can be transferred to $q_k$ given in~\eqref{009} (see the following subsection~C for more explanation). For solving~\eqref{017}, we introduce two assumptions: 1) the~\emph{a priori} error and input signals are uncorrelated, i.e., the orthogonality principle~\cite{sayed2003fundamentals}; 2) the forgetting factor is deterministic at each time~\cite{paleologu2008robust,paleologu2014practical}. As a result, the solution of~\eqref{017} with respect to $\lambda_k$ is given as
\begin{equation}
\label{018}
\begin{array}{rcl}
\begin{aligned}
\lambda_k = \frac{\sigma_{\theta,k} \sigma_v}{\sigma_{e,k} - \sigma_v},
\end{aligned}
\end{array}
\end{equation}
where $\sigma_{e,k}^2 = E\{e_k^2\}$ and $\sigma_{\theta,k}^2=E\{\theta_k^2\}$ are the variances of the corresponding signals. Both $\sigma_{e,k}^2$ and $\sigma_{\theta,k}^2$ could be estimated in a recursive way:
\begin{subequations} \label{eq:19}
    \begin{align}
    %   \begin{array}{rcl}
    %   \begin{aligned}
    \hat{\sigma}_{e,k}^2 &= \chi \hat{\sigma}_{e,k-1}^2 + (1-\chi) q_k^2 e_k^2,
    %    \end{aligned}
    %   \end{array}
    \label{eq:19a}\\
    \hat{\sigma}_{\theta,k}^2 &= \chi \hat{\sigma}_{\theta,k-1}^2 + (1-\chi) \theta_k^2, \label{eq:19b}
    \end{align}
\end{subequations}
where $q_k^2$ aims to reduce the negative influence of the impulsive noise on the estimated variance $\hat{\sigma}_{e,k}^2$ and $\chi\in[0.9, 1)$ is a the smoothing factor. To estimate the variance $\sigma_{v}^2$ of the background noise, we extend the approach in~\cite{paleologu2014practical} to impulsive noise environments, formulated as
\begin{equation}
\label{020}
\begin{array}{rcl}
\begin{aligned}
\hat{\sigma}_{v,k}^2 = \frac{\hat{\sigma}_{d,k}^2 \hat{\sigma}_{e,k}^2}{\hat{\sigma}_{e,k}^2 +\hat{\sigma}_{y,k}^2},
\end{aligned}
\end{array}
\end{equation}
where the estimated powers $\hat{\sigma}_{d,k}^2$ and $\hat{\sigma}_{y,k}^2$ are calculated similar to~\eqref{eq:19a} as
\begin{subequations} \label{eq:21}
    \begin{align}
    %   \begin{array}{rcl}
    %   \begin{aligned}
    \hat{\sigma}_{d,k}^2 &= \chi \hat{\sigma}_{d,k-1}^2 + (1-\chi) q_k^2 d_k^2,
    %    \end{aligned}
    %   \end{array}
    \label{eq:21a}\\
    \hat{\sigma}_{y,k}^2 &= \chi \hat{\sigma}_{y,k-1}^2 + (1-\chi) y_k^2, \label{eq:21b}
    \end{align}
\end{subequations}
with $y_k = \bm x_{k}^\text{T}\bm w_{k-1}$ being the output of the adaptive filter. Due to using the power estimates, it could happen that $\hat{\sigma}_{e,k}^2 < \hat{\sigma}_{v,k}^2$; however,~\eqref{018} shows that this situation has to be prevented. As such, we could set $\lambda_k$ to $\lambda_{\max}$ which is close to one. Furthermore, in the steady-state $\hat{\sigma}_{e,k}^2$ may vary around $\hat{\sigma}_{v,k}^2$. Therefore, we propose a more practical solution for~$\lambda_k$ by imposing the convergence state $\hat{\sigma}_{e,k}^2 < \tau \hat{\sigma}_{v,k}^2$ with $\tau\in [1, 2]$, as follows:
\begin{equation}
\label{022}
\begin{aligned}
\lambda_k =  \left\{ \begin{aligned}
&\lambda_{\max},\;\text{if } \hat{\sigma}_{e,k}^2 < \tau \hat{\sigma}_{v,k}^2,\\
&\min \left\lbrace  \frac{\hat{\sigma}_{\theta,k} \hat{\sigma}_v}{|\hat{\sigma}_{e,k} - \hat{\sigma}_v|+\kappa}, \lambda_{\max} \right\rbrace,\; \text{otherwise},
\end{aligned} \right.
\end{aligned}
\end{equation}
where $\kappa$ is a small positive constant. This completes the VFF's derivation for the S-RRLS algorithm.

By equipping S-RRLS with the proposed $\hat{\rho}_k^{\text{opt}}$ and $\lambda_k$ adaptations, we arrive at the JO-S-RRLS algorithm. In fact, the proposed~$\hat{\rho}_k^{\text{opt}}$ and $\lambda_k$ originate from the alternating optimization idea which is a powerful way to solve the challenging global optimization problems~\cite{hong2016convergence,de2014sparsity}.

Remark~1: As can be seen in~\eqref{012}, the term $\bigtriangleup_k$ results from the sparsity-aware step~\eqref{eq:10b}. Only when $\bigtriangleup_k<0$, the S-RRLS algorithm will work better than the RRLS algorithm. Accordingly, $\rho$ must satisfy the inequality $0<\rho<2\rho_k^{\text{opt}}$. Following a similar derivation in Appendix~D in \cite{yu2021sparsity}, $\bigtriangleup_k<0$ is likely to be true when identifying a sparse vector~$\bm w^o$. It follows that the optimal $\rho_k^{\text{opt}}$ given in~\eqref{014} may exist. In the future, we will analyze the mean and mean-square behaviors of the JO-S-RRLS algorithm.
\subsection{Practical considerations}

For implementing the S-RRLS and JO-S-RRLS algorithms, two problems should be addressed.

Firstly, the robustness of the algorithms against impulsive noises relies on how to design the robustness function $\varphi(e_k)$ to further obtain $q_k$ in~\eqref{009}. As an example, the modified M-estimator is used~\cite{chan2004recursive}, i.e., $\varphi(e_k) = \left\{ \begin{aligned} &e_k^2/2, \text{ if } |e_k| \leq \xi,\\ &\xi^2/2, \text{ if } |e_k| > \xi, \end{aligned} \right.$ such that $ q_k = \left\{ \begin{aligned} &1, \text{ if } |e_k| \leq \xi,\\&0, \text{ if } |e_k| > \xi, \end{aligned} \right.$. It reveals that when $|e_k| > \xi$ holds (generally, when the impulsive noise happens), the Kalman gain $\bm K_k$ will be a zero vector due to $q_k=0$, thus stopping the update of the adaptive filter. The threshold $\xi$ is chosen as $\xi =\vartheta \hat{\sigma}_{\varepsilon,k}$,\; $\hat{\sigma}_{\varepsilon,k}^2 = \zeta \hat{\sigma}_{\varepsilon,k-1}^2 + c_\sigma (1-\zeta) \text{med}(\bm a_k^\varepsilon)$, where $\zeta \in [0.9,1)$ is an exponentially weighting factor (except $\zeta=0$ at time $k=0$), the median operator $\text{med}(\cdot)$ is to remove data in the window $\bm a_k^\varepsilon=[e_k^2,e_{k-1}^2,...,e_{k-N_w+1}^2]$ disturbed by impulsive noises, and $c_\sigma = 1.483(1+5/(N_w-1))$ is the correction factor. Note that the window length~$N_w$ needs to be properly chosen. Larger $N_w$ provide a more robust estimate $\hat{\varepsilon}_{e,k}^2$ but require a higher complexity~\cite{chan2004recursive}. Also, the value of $\vartheta$ is often chosen as 2.576~\cite{chan2004recursive}.
%It means that there is a 99\% confidence to prevent $e_k$ from contributing to the update for $|e_k| \geq \xi$ under the assumption that $e_k$ is assumed to be Gaussian (which is reasonable except when being polluted by impulsive noise).

Secondly, to effectively characterize the sparsity of systems, how to design $f(\cdot)$ in~\eqref{002} is also a key factor. Here we choose the popular log-penalty~$f(\bm w_k) = \sum_{m=1}^{M} \ln(1+|w_{m,k}|/\mu)$~\cite{yu2021sparsity}, where $w_{m,k}$ denotes the $m$-th entry of $\bm w_k$, and $\mu>0$ denotes the shrinkage factor that helps to distinguish non-zero and zero entries. Thus, the entries of $\triangledown (\bm \psi_{m,k})$ in~\eqref{eq:10b} are given by~$\triangledown (\psi_{m,k}) = \frac{\text{sgn}(\psi_{m,k})}{\mu + |\psi_{m,k}|}, m=1,...,M$.

It needs to point out that other robust criteria ~\cite{navia2012combination,zayyani2014continuous,lu2018recursive,lu2016improved,radmanesh2018recursive,roth2004generalized} and sparsity-aware penalties~\cite{eksioglu2011rls,zakharov2013dcd,yu2021sparsity,de2014sparsity} can also be applied to present different JO-S-RRLS algorithms. However, discussing the effects of different choices of $\varphi(e_k)$ or $f(\bm w_k)$ is not the focus of this paper.

Remark~2: Compared with the original S-RRLS algorithm, the extra computational complexity of the JO-S-RRLS algorithm stems from the adaptations of $\rho_k$ and $\lambda_k$, which requires $M^2+3M+13$ multiplications, $M^2+2M+6$ additions, $3$ divisions, and $3$ square-roots per iteration.
%\subsection{Computational Complexity}
%The direct solution of~\eqref{003} is $\bm w_n = \bm R_n^{-1}\bm z_n$. The regularization $\delta_n$ is to maintain the numerical stability of this solution~\cite{sayed2003fundamentals}. However, this leads to the complexity of $\mathcal{O}(M^3)$ due to the matrix inversion $\bm R_n^{-1}$. Generally, $\delta_n$ is chosen as $\delta_n=\lambda^{n+1}\delta_0$ (e.g., in this paper), it makes \eqref{004} become $\bm R_n = \lambda \bm R_{n-1}+f_n \bm x_n \bm x_n^\text{T}$. Then, using the matrix inversion lemma, $\bm R_n^{-1}$ can be calculated in a recursive way so that the complexity of the resulting algorithm is $\mathcal{O}(M^2)$, while it is still high for large $M$.
%\begin{table}[tbp]
%   \scriptsize
%   \centering
%   \caption{Complexity of Algorithms per Input Sample}
%   \label{table_4}
%   \begin{tabular}{@{}l|c|c|cc}
%       \hline
%       \text{Algorithms} &\text{Additions} &\text{Multiplications} &\text{Divisions}\\
%       \hline
%       \text {LMS}  &$2M$ &$2M+1$ &0\\ \hline
%       \text {(R) RLS}  &$3M^2+M$ &$4M^2+4M+1$ &1\\ \hline
%       \begin{tabular}[c]{@{}l@{}} \text {(R) DCD RLS}\\ (general input)\end{tabular} &$M^2+2M+P_a$ &$\frac{3}{2}M^2+\frac{7}{2}M+1$ &0\\ \hline
%       \begin{tabular}[c]{@{}l@{}} \text {(R) DCD RLS}\\ (tap-delayed input)\end{tabular}   &$3M+P_a$ &$5M+2$ &0\\
%       \hline
%   \end{tabular}
%\end{table}
\section{Simulation Results}
In this section, simulations are presented to evaluate the proposed JO-S-RRLS algorithm. It is assumed that the length of the adaptive filter is the same as that of the unknown sparse vector $\bm w^o$. The input signal $x_k$ is generated by filtering a zero-mean white Gaussian random process $\nu_k$ with unit variance through a second-order autoregressive model $x_k = 0.4 x_{k-1}- 0.4 x_{k-2} + \nu_k$. The normalized mean square deviation is used as the performance measure, defined as $\text{NMSD}_k= 10\log_{10}(||\bm w_k -\bm w^o||_2^2/||\bm w^o||_2^2)$. All the curves are obtained by averaging the results over 100 independent runs.

\begin{figure}[htb]
    \centering
    \includegraphics[scale=0.52] {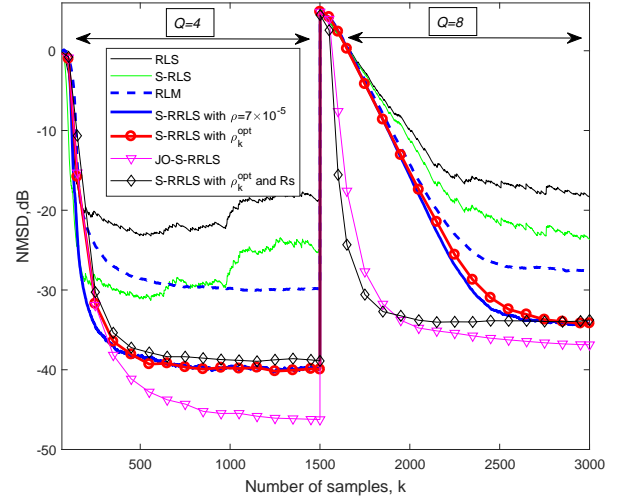}
%\\
%   \includegraphics[scale=0.45] {fig1b.eps}
    \vspace{-1em} \caption{NMSD performance of RLS-type algorithms in the CG noise with~$p=0.001$. The VFF's parameters are set to $\lambda_{\max}=0.99999,\chi=0.96$, and $\tau=1.5$. }
    \label{Fig1}
 \end{figure}

Case~1: The sparse vector~$\bm w^o$ has $M=64$ elements, and its non-zero elements follow from a zero-mean Gaussian distribution with variance~$1/\sqrt{Q}$ and their positions are randomly selected from the binomial distribution, with $Q$ being the number of non-zero elements. A smaller $Q$ means sparser $\bm w^o$. The vector~$\bm w^o$ cardinality changes from $Q=4$ to $Q=8$ at time~$k=1501$. The noise with impulsive behavior $v_k$ is drawn from the contaminated-Gaussian (CG) process, i.e., $v_k=v_k^\text{g} + v_k^\text{i}$. Specifically, $v_k^\text{g}$ is zero-mean white Gaussian noise, with variance $\sigma_\text{g}^2$ given by the signal-to-noise ratio of 30 dB. The impulsive noise $v_k^\text{i}$ is described as $v_k^\text{i} =b_k \eta_k$, where $b_k$ follows from the Bernoulli distribution with the probabilities $P\{b_k=1\}=p$ and $P\{b_k=0\}=1-p$, and $\eta_k$ is also zero-mean white Gaussian noise with a large variance of $\sigma_\eta^2 =1000 E\{(\bm x_k^\text{T} \bm w^o)^2\}$. Fig.~\ref{Fig1} compares the proposed S-RRLS and JO-S-RRLS algorithms with the existing RLS, S-RLS, and RLM algorithms. To fairly evaluate them, we set $\lambda=0.995$ and $\delta=0.5$ for all the algorithms, the M-estimate parameter $N_w=9$ and $\zeta=0.99$ for all the robust algorithms, the log-penalty parameter $\mu=0.01$ for all the sparsity-aware algorithms. As expected, the RLS and S-RLS algorithms are not suitable for impulsive noise scenarios due to the performance degradation, while other algorithms show good robustness. Benefited from the sparsity-aware step, S-RRLS algorithm reduces the steady-state misadjustment in sparse systems as compared to the RLM algorithm. Also, by using the proposed adaptation of $\rho$, the S-RRLS algorithm with $\rho_k^\text{opt}$ avoids the choice problem of $\rho$. To deal with the sudden change of $\bm w^o$, we may also resort to the reset (Rs) technique\footnote{Here the Rs technique modifies the one in~\cite{shi2019combined} due to the presence of impulsive noises, i.e., if $\log(q_k^2e_k^2/e_{\text{avr},k-1}^2)>1.5$, we reset $\bm P_k$ with $\bm P_0$, where $e_{\text{avr},k-1}^2 = t e_{\text{avr},k-1}^2 + (1-t)q_k^2e_k^2$ with $t=0.98$.}, namely, the S-RRLS with $\rho_k^\text{opt}$ and Rs algorithm recovers the tracking capability. Importantly, by jointly optimizing~$\rho$ and $\lambda$, the proposed JO-S-RRLS algorithm not only obtains a low steady-state misadjustment but also good tracking capability.

\begin{figure}[htb]
    \centering
    \includegraphics[scale=0.52] {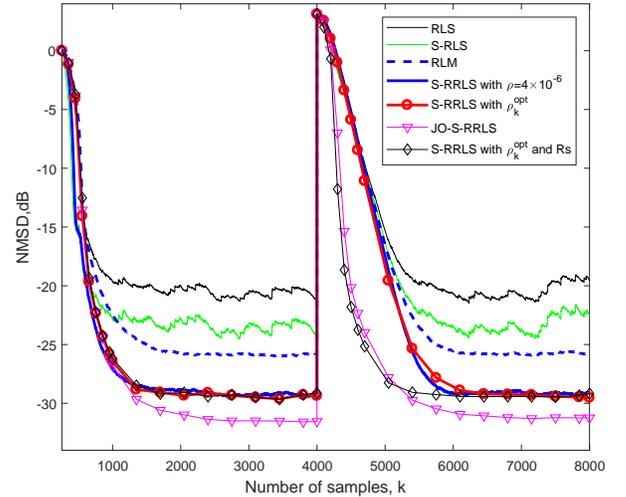}
    \hspace{2cm}\caption{NMSD performance of RLS-type algorithms in the $\alpha$-stable noise. Some parameters are retuned as: $\lambda=0.977$, $N_w=16$, and $\mu=0.001$ for all the algorithms.}
    \label{Fig2}
\end{figure}

Case 2: The sparse vector~$\bm w^o$ is the echo channel~1 from the ITUT G.168 standard, with $M = 256$ taps~\cite{stnec2015}. The noise with impulsive behavior $v_k$ follows from the symmetric $\alpha$-stable random process, called the $\alpha$-stable noise. Its characteristic function~\cite{nikias1995signal} is given as $\phi(t)=\exp(-\gamma \lvert t \lvert^\alpha)$, where $\alpha \in (0,2]$ describes the impulsiveness of the noise (smaller $\alpha$ corresponds to stronger impulsive noises) and $\gamma>0$ is similar to the variance of the noise. Note that when $\alpha=2$, it reduces to the Gaussian distribution. In Fig~\ref{Fig2}, we set $\alpha=1.65$ and $\gamma=0.02$. As one can see, the proposed JO-S-RRLS algorithm achieves the best performance among these algorithms, since it optimizes the parameters $\rho$ and $\lambda$ simultaneously.
\section{Conclusion}
In this work, a unified S-RRLS algorithm update was presented, which aims at identifying sparse systems in impulsive noise. By replacing straightforwardly the specified robustness criterion and sparsity-aware penalty, it can lead to different S-RRLS algorithms. We then developed adaptive schemes for both the forgetting factor and the sparsity penalty parameter in the S-RRLS algorithm, thus arriving at the JO-S-RRLS algorithm with a further improved performance in terms of the steady-state misadjustment and tracking capability. Simulations in various impulsive noise scenarios have been conducted to verify the effectiveness of the proposed algorithms.

%\appendices
%\numberwithin{equation}{section}
%\section{Derivation of \eqref{060}}
%\renewcommand{\theequation}{\thesection.\arabic{equation}}

\ifCLASSOPTIONcaptionsoff
  \newpage
\fi

\bibliographystyle{IEEEtran}
\bibliography{IEEEabrv,mybibfile}

%\begin{thebibliography}{1}

%\bibitem{IEEEhowto:kopka}

%H.~Kopka and P.~W. Daly, \emph{A Guide to \LaTeX}, 3rd~ed.\hskip 1em plus
% 0.5em minus 0.4em\relax Harlow, England: Addison-Wesley, 1999.

%\end{thebibliography}

% biography section
%
% If you have an EPS/PDF photo (graphicx package needed) extra braces are
% needed around the contents of the optional argument to biography to prevent
% the LaTeX parser from getting confused when it sees the complicated
% \includegraphics command within an optional argument. (You could create
% your own custom macro containing the \includegraphics command to make things
% simpler here.)
%\begin{IEEEbiography}[{\includegraphics[width=1in,height=1.25in,clip,keepaspectratio]{mshell}}]{Michael Shell}
% or if you just want to reserve a space for a photo:

%\begin{IEEEbiography}{Michael Shell}
%Biography text here.
%\end{IEEEbiography}

% if you will not have a photo at all:
%\begin{IEEEbiographynophoto}{John Doe}
%Biography text here.
%\end{IEEEbiographynophoto}

% insert where needed to balance the two columns on the last page with
% biographies
%\newpage

%\begin{IEEEbiographynophoto}{Jane Doe}
%Biography text here.
%\end{IEEEbiographynophoto}

% You can push biographies down or up by placing
% a \vfill before or after them. The appropriate
% use of \vfill depends on what kind of text is
% on the last page and whether or not the columns
% are being equalized.

%\vfill

% Can be used to pull up biographies so that the bottom of the last one
% is flush with the other column.
%\enlargethispage{-5in}

\end{document}